\documentclass[showpacs,preprintnumbers,10pt,onecolumn]{revtex4}%
\usepackage{amssymb}
\usepackage{amsfonts}
\usepackage{amsmath}
\usepackage{graphicx}
\usepackage{times}
\usepackage{dcolumn}
\usepackage{bm}
\usepackage{revsymb}
\usepackage{color}%

\begin{document}
\title{Impulse-induced localized nonlinear modes in an electrical lattice}
\author{Faustino Palmero $^{1}$, Jes\'{u}s Cuevas-Maraver$^{2}$, Lars Q. English$^{3}$,
and Ricardo Chac\'{o}n $^{4}$}
\affiliation{$^{1}$Grupo de F\'{\i}sica No Lineal, Departamento de F\'{\i}sica Aplicada I,
Escuela T\'{e}cnica Superior de Ingenier\'{\i}a Inform\'{a}tica, Universidad
de Sevilla, Avda Reina Mercedes s/n, E-41012 Sevilla, Spain}
\affiliation{$^{2}$Grupo de F\'{\i}sica No Lineal, Departamento de F\'{\i}sica Aplicada I,
Escuela Polit\'{e}cnica Superior, Universidad de Sevilla, Virgen de \'{A}frica
7, 41011 Sevilla, Spain and Instituto de Matem\'{a}ticas de la Universidad de
Sevilla (IMUS), Edificio Celestino Mutis, Avda Reina Mercedes s/n, E-41012
Sevilla, Spain}
\affiliation{$^{3}$Department of Physics and Astronomy Dickinson College, Carlisle,
Pennsylvania, 17013, USA}
\affiliation{$^{4}$Departamento de F\'{\i}sica Aplicada, E.I.I., Universidad de
Extremadura, Apartado Postal 382, E-06006 Badajoz, Spain and Instituto de
Computaci\'{o}n Cient\'{\i}fica Avanzada (ICCAEx), Universidad de Extremadura,
E-06006 Badajoz, Spain}
\date{\today}

\begin{abstract}
Intrinsic localized modes, also called discrete breathers, can exist under
certain conditions in one-dimensional nonlinear electrical lattices driven by
external harmonic excitations. In this work, we have studied experimentally the efectiveness of generic
periodic excitations of variable waveform at generating discrete breathers in
such lattices. We have found that this generation phenomenon is optimally
controlled by the impulse transmitted by the external excitation (time
integral over two consecutive zeros), \textit{irrespectively} of its particular
waveform.
\end{abstract}
\pacs{05.45.Xt, 05.45.Pq, 87.18.Bb, 74.81.Fa}
\maketitle

\section{Introduction} Intrinsic localized modes, or
discrete breathers (DBs), can exist in a wide variety of coupled nonlinear
oscillator networks under very general conditions \cite{breather_reviews,
Aubry}. Specifically, they have been experimentally observed in periodically driven
dissipative systems, such as Josephson junction arrays \cite{JJ}, coupled
pendula chains \cite{pendula}, micro- and macro-mechanical cantilever arrays
\cite{cantilever}, granular crystals \cite{granular}, and nonlinear electrical
lattices \cite{electric1, electric2}.

In all these cases, the external periodic excitations (PEs) have systematically been
taken as harmonic excitations. However, there exists a vast
diversity of nonlinear PEs depending upon the particular physical context
under consideration. The relevance of the excitation waveform,
which reflects the spectral content of the excitation's Fourier expansion, has
previously been pointed out in many different backgrounds, such as ratchet
transport \cite{2}, adiabatically ac driven periodic (Hamiltonian) systems
\cite{3}, driven two-level systems and periodically curved waveguide arrays
\cite{4}, chaotic dynamics of a pump-modulation Nd:YVO$_{4}$ laser \cite{5},
topological amplification effects in scale-free networks of signaling devices
\cite{6}, and controlling chaos in starlike networks of dissipative nonlinear
oscillators \cite{7}. In all these previous works, the external $T$-PE $F(t)$
is chosen as a generic periodic function of zero-mean having equidistant
zeros, for which it has been shown both theoretically and numerically that the
\textit{impulse} transmitted over a half period, $I=\int_{0}^{T/2}F(t)dt$, is
the relevant quantity characterizing its clear-cut dynamical effect. It is
worth noting that the relevance of the excitation impulse comes ultimately
from the fact that it takes into account the \textit{conjoint} effects of its
amplitude, period, and waveform, on the one hand, and from the existence of a
correlation between variations of impulse and subsequent variations of the
energy transmitted by the PE, on the other hand.

Regarding DBs, it has recently been shown that the generation of stationary
and moving DBs appearing in prototypical nonlinear oscillator networks
subjected to non-harmonic PEs are optimally controlled by solely varying the
impulse transmitted by the PEs, while keeping constant their amplitude and
period \cite{CCP16}. Motivated by these results, we will focus in the present
work on the experimental generation of stationary DBs in a nonlinear
electrical lattice driven by non-harmonic PEs. We will demonstrate
experimentally that this generation phenomenon is optimally controlled by the impulse transmitted by
the PEs, irrespective of its particular waveform.

\section{Experimental and theoretical setup}
\begin{figure}[ptb]
\includegraphics[width=0.8\textwidth]{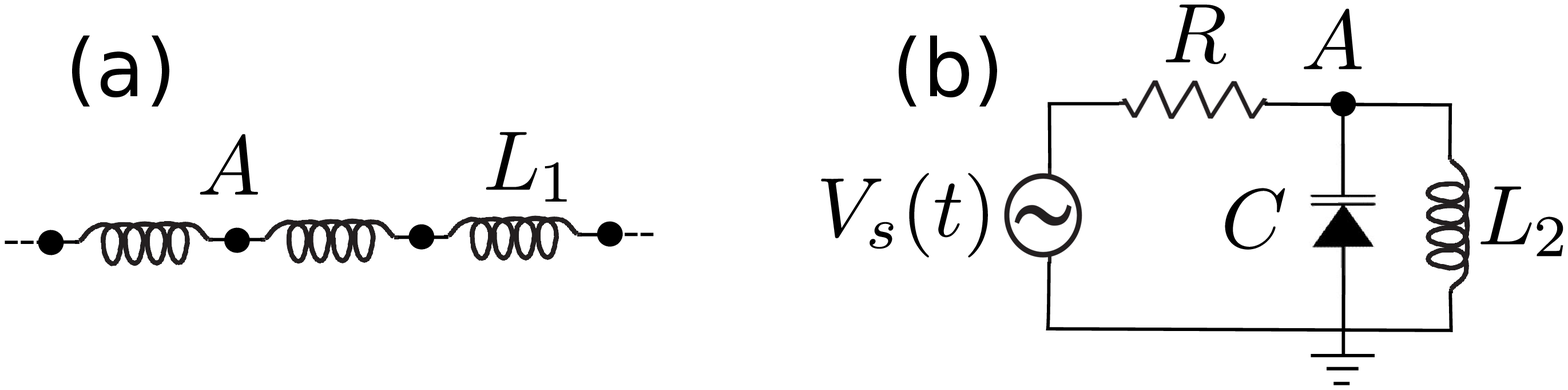}
\caption{Schematic circuit
diagrams of the electrical transmission line (a), where the black points
represent circuit cells (b). Each cell is connected to a
periodic voltage source $V_{s}(t)$ via a resistor $R$, and grounded. Each
point $A$ of an elemental circuit is connected via inductors $L_{1}$ to the
corresponding points $A$ of neighboring cell. Voltages are
monitorized at point $A$.}%
\label{lattice}%
\end{figure}

Our system, shown in Fig. \ref{lattice}, is a
simple electrical line previously considered in Ref. \cite{electric1}. Its connected
nodes (circuits cells made of inductors $L_{2}$ and load resistances $R$)
become nonlinear when including a varactor diode (NTE 618) in each of them, being the
last circuit cell connected to the first one (periodic boundary
conditions). The periodic driving is chosen to be spatially homogeneous; to this aim, we excite the electrical line with any
previously chosen $T$-periodic function $V_{s}(t)$ by means of a programmable waveform generator. The voltage $V_n$ at each lattice node $n$ at point $A$
can be readily monitorized by using oscilloscopes. In our lattice $L_1=0.68$ mH, $L_2= 0.33$ mH, both with a tolerance of 10\%, $R=10$ k$\Omega$ with a tolerance of 5\% and the number of nodes $N=10$; although it may look a small number of nodes, it is large enough to get 
isolated one-peak DBs \cite{electric1}.

As explicitly shown below, we will focus on the case corresponding to an
external voltage source in the form $V_{s}(t)=A_{s}g(t)$, where $g(t)$ is a
dimensionless and conveniently normalized $T$-periodic function and $A_{s}$ is a voltage.

In the absence of dissipation and external PEs, an isolated 
circuit cell becomes a nonlinear capacitance $C(V)$ associated in parallel to an inductor $L_2$, a Hamiltonian system where
\begin{equation}
H(p,q)=\frac{p^2}{L_2C(q)^2}+\int_{0}^{q} x C(x) \mathrm{d}x,
\end{equation}
$p=L_2C^{2}\dot{V}$ and  $q=V$. In the presence of the resistance $R$ and the external PE, the variation of the
energy is written as
\begin{align}
\frac{dE}{dt} &  =L_2 C(V)\frac{dV}{dt}\left[\frac{1}{R}%
\frac{d V_s(t)}{d t}-\left(  \frac{dI_{d}(V)}{dV}+\frac{1}{R}\right)
\frac{d V}{d t}\right].
\end{align}

Integrating over half  period $T$ (once the system has reached a steady state) and applying
the first mean value theorem for integrals \cite{6,7}, one straightforwardly
obtains
\begin{eqnarray}
\Delta E  & = & -\left[\frac{1}{C}\left(  \frac{1}{R}+\frac{dI_{d}}{dV}\right)  \right]  _{t=t^{\ast}} \int_{T/2} p \mathrm{d} q%
+\nonumber\label{deltaE}\\
&  & \left[  \frac{p}{RC}\right]  _{t=t^{\ast
\ast}}I,
\end{eqnarray}
where $t^{\ast}$, $t^{\ast\ast}$ $\in\lbrack0,T/2]$. According to \cite{CCP16}, the role  of the external periodic excitation $F(t)$ is played by the explicit time--dependent
function $dV_s(t)/dt$
, and hence $I=\int_{0}^{T/2}F(t)dt=V_s(T/2)-V_s(0)$ is the impulse. Note that in a more accurate model with additional dissipation
sources, the impulse contribution will be unchanged.

In some cases, the basic dynamics of the electrical
network composed by the aforementioned circuit cells coupled by inductors
$L_{1}$ can be qualitatively described by a simple model as proposed in \cite{electric1, electric2}, where the introduction of a phenomenological
resistor in the model is enough to reproduce some experimental results. Nevertheless, in general,  a more sophisticated
model is necesssary to match experiments. In this paper we will focus only
on experimental data and the formulation of an accurate model will be object of further work.

In order to study the impulse-induced
generation of stationary DBs, we conveniently chose the periodic function
$dV_s/d t$ as given in terms of Jacobian elliptic functions. Indeed, after
normalizing their (natural) arguments to keep their period as a fixed
independent parameter, their waveforms can be suitably changed by solely
varying a single parameter: the (elliptic) shape parameter $m$, and hence the
corresponding impulse will only depend on $m$ once the amplitude and the
period are fixed. To demonstrate that our results are independent of the
particular selection of the PE, we considered two different choices:
\begin{align}
\frac{dV_{s}^{(1)}(t)}{dt } &  =\frac{A_s}{T}\medspace\text{sn}\left[
\frac{4K(m)}{T}t;m\right]  ,\label{g1}\\
\frac{dV_{s}^{(2)}(t)}{d t} &  =\frac{A_s}{T}N(m)\medspace\text{sn}%
\left[  \frac{4K(m)}{T}t;m\right]  \times\nonumber\\
&  \text{dn}\left[  \frac{4K(m)}{T} t;m\right]  ,\label{g2}%
\end{align}
where $N(m)$ is a normalization factor \cite{CCP16}, $K(m)$ is the complete
elliptic integral of the first kind, and $\operatorname{sn}\left(
\cdot;m\right)  ,\operatorname*{dn}\left(  \cdot;m\right)  $ are Jacobian
elliptic functions of parameter $m$. Thus, one can change the PEs' waveform by
solely varying their shape parameter $m$ between 0 and 1 while keeping
constant their amplitude and period. If parameter $m$ is small enough ($m \in [0,0.8]$), a good approximation of these functions is given by
the first two terms of its Fourier series which reads
\begin{align}
\frac{dV_{s}^{(1,2)}(t)}{dt } &  =\frac{A_s}{T}\left[G^{(1,2)}_1(m) \sin\left(\frac{2 \pi}{T} t\right)+ \right. \nonumber \\ & \left. G^{(1,2)}_3 (m) \sin\left(\frac{6 \pi}{T} t\right)\right],\label{g1a}
\end{align}
being 
\begin{align}
G^{(1)}_1 & =\frac{2 \pi}{\sqrt{m} K(m)} \frac{q^{1/2}}{1-q}, \\
G^{(1)}_3 & =\frac{q(1-q)}{1-q^3} G^{(1)}_1, \\
G^{(2)}_1 & =N(m) \frac{\pi^2}{\sqrt{m} K^2(m)} \frac{q^{1/2}}{1+q}, \\
G^{(2)}_3 & =\frac{3q(1+q)}{1+q^3 } G^{(2)}_1, \\
\end{align}
and $q=\exp{(-\pi K(1-m)/K(m))}$.  In Fig. \ref{fourier} we depict the normalized second harmonic,
$G_{3}(m)/G_{1}(m)$ as functions of the shape parameter $m$, for the PEs (\ref{g1}) and (\ref{g2}).

\begin{figure}[ptb]
\includegraphics[width=0.4\textwidth]{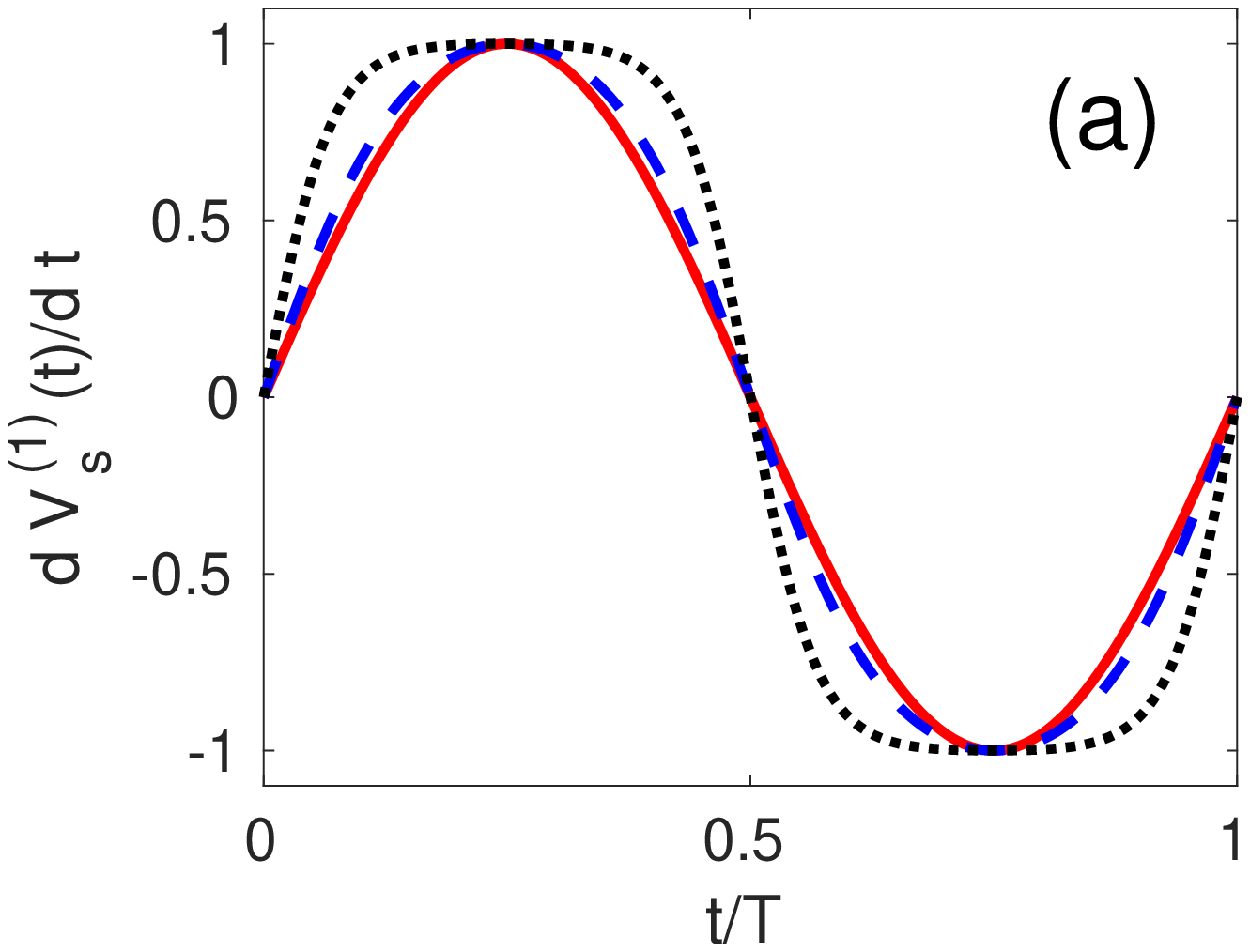}
\includegraphics[width=0.4\textwidth]{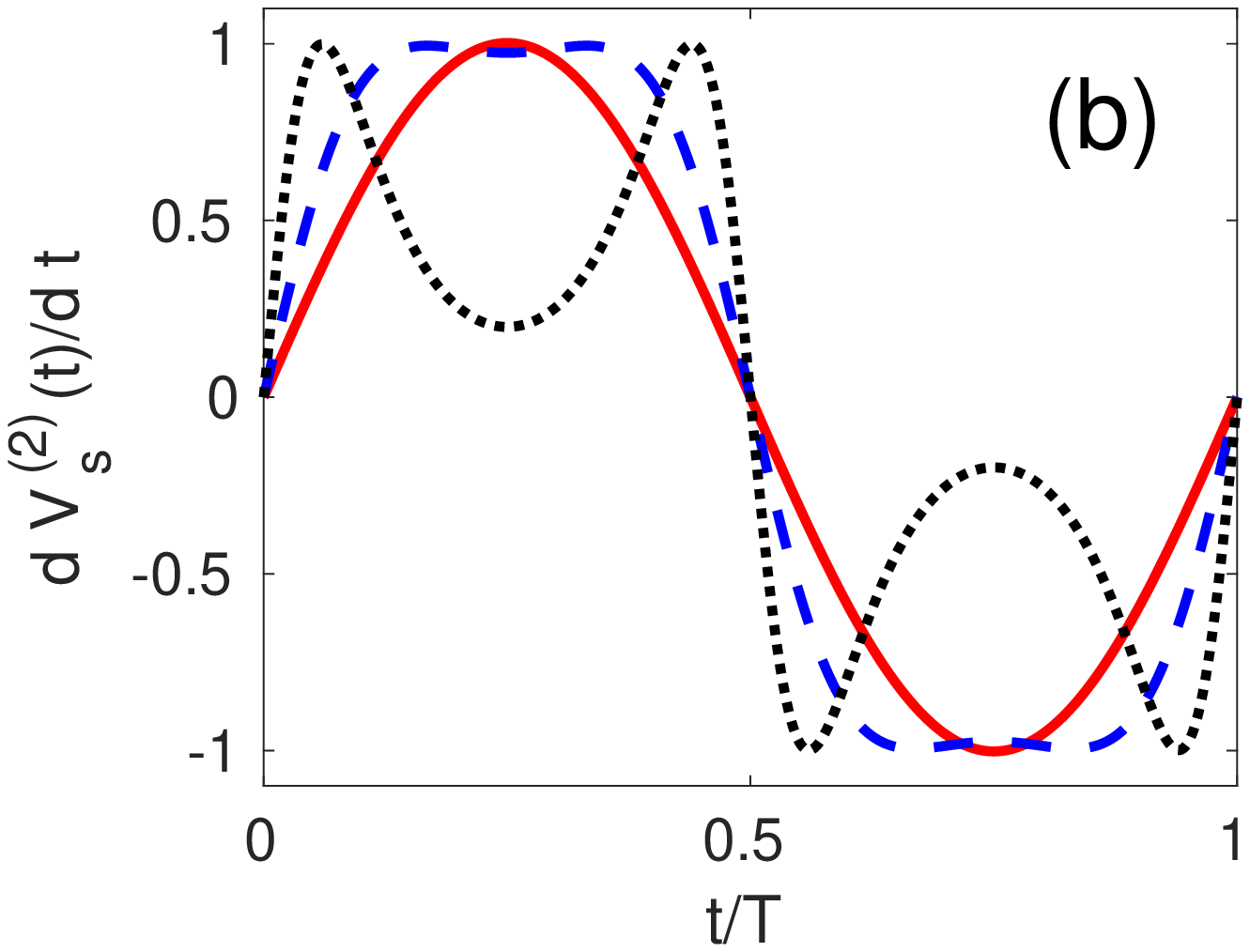}
\caption{(Color online)
Normalized periodic excitations [given by (\ref{g1}) in panel (a) and by (\ref{g2}) in panel (b)]
vs time over
a period $T$ for three values of the shape parameter: $m=0$
(red solid line), $m=0.6$ (blue dashed
line), and $m=0.99$ (black dotted line)}.%
\label{driving}%
\end{figure}

\begin{figure}[h]
\includegraphics[width=0.4\textwidth]{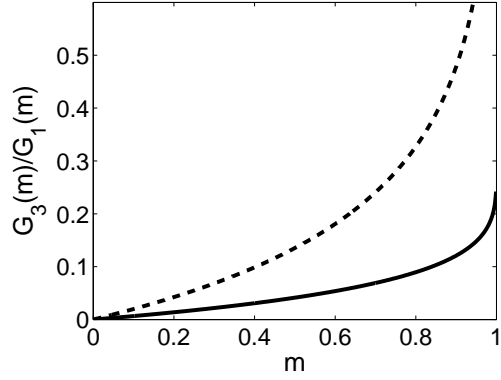}
\caption{(Color online) Normalized Fourier coefficients of the PEs (\ref{vs1}) and (\ref{vs2}) vs
shape parameter $m$: $G^1_{3}(m)/G^1_{1}(m)$ of the PE (\ref{vs1}) (black solid line) and
$G^2_{3}(m)/G^2_{1}(m)$ of the PE (\ref{vs2}) (black dash-dot line).}%
\label{fourier}%
\end{figure}

\begin{figure}[ptb]
\includegraphics[width=0.4\textwidth]{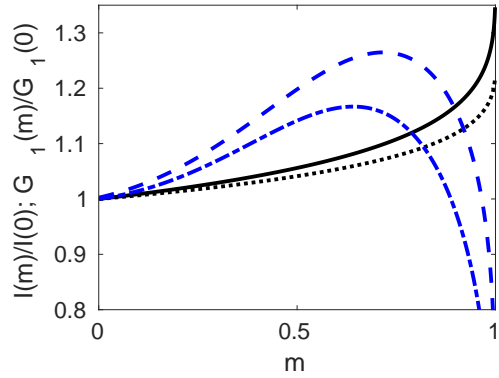}\caption{(Color online)
Normalized impulse functions $I^{(1,2)}(m)/I^{(1,2)}(m=0)$
and first Fourier coefficient $G^{(1,2)}_1(m)/G^{(1,2)}_1(m=0)$ as
functions of the shape parameter $m$. Solid (black) and dashed (blue) lines
correspond to the functions (\ref{I1}) and (\ref{I2}) while
dotted (black) and dash-dot (blue) lines correspond to the
first Fourier coefficient of (\ref{g1}) and (\ref{g2}), respectively.}%
\label{impulso}%
\end{figure}

The corresponding driving functions $V_{s}^{\left(  1,2\right)
}(t)$ are written as%

\begin{align}
V_{s}^{(1)}(t) &  =\frac{A_{s}}{4\sqrt{m}K(m)}\left\{  \ln\left[
\text{dn}\left(  \frac{4K(m)}{T}t;m\right)  \right.  \right.  -\nonumber\\
&  \left.  \left.  \sqrt{m}\medspace\text{cn}\left(  \frac{4K(m)}%
{T}t;m\right)  \right]  -\right.  \nonumber\\
&  \left.  \ln\left(  \sqrt{1-m}\right)  \right.  \bigg\},\label{vs1}\\
V_{s}^{(2)}(t) &  =-\frac{A_{s}N(m)}{4K(m)}\text{cn}\left(  \frac{4K(m)}%
{T}t;m\right)  ,\label{vs2}%
\end{align}
where $\text{cn}(\cdot;m)$ is the Jacobian elliptic function of parameter $m$,
and hence they are shift-symmetric functions: $V_{s}^{(1,2)}(t)=-V_{s}%
^{(1,2)}(t+T/2)$. The corresponding impulse functions read

\begin{align}
I^{(1)}(m) &  =\frac{A_{s}}{4\sqrt{m}K(m)}\ln\left(  \frac{1+\sqrt{m}}%
{1-\sqrt{m}}\right)  ,\label{I1}\\
I^{(2)}(m) &  =\frac{A_{s}N(m)}{2K(m)},\label{I2}%
\end{align}
respectively. Figure \ref{impulso} shows plots of the (normalized) functions
$I^{(1,2)}(m)$ over the entire range $m\in\left[  0,1\right]  $. It is worth
noticing that these functions present different properties: While the impulse
function $I^{(1)}\left(  m\right)  $ presents a monotonically increasing behavior
for every value of $m$, the impulse function $I^{(2)}\left(
m\right)  $ presents a single maximum at $m=m_{max}\simeq0.717$. Although the
aforementioned values $t^{\ast},t^{\ast\ast}$ will generally depend upon
the shape parameter (see Eq.(\ref{deltaE})), they become independent of the
PE's waveform as $T\rightarrow0$ \cite{6,7}. Note, however, that this is an
unreachable limit owing to the DBs frequencies are necessarily below a certain
threshold value.
The first Fourier coefficient, $G^{(1,2)}_1$, of the PEs (\ref{g1}) and
(\ref{g2}) is also represented in Fig. \ref{impulso}, featuring a qualitatively similar behaviour to that of the impulse.

\

\section{Impulse-induced DBs scenario} 
\begin{figure}[h]
\includegraphics[width=0.5\textwidth]{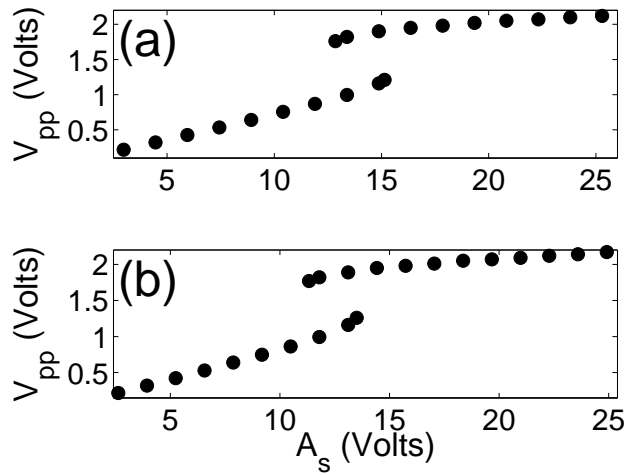}
\caption{(Color online) Response of an
isolated circuit cell $V_{pp}$ (peak-to-peak value of $V(t)$) as a
function of the amplitude $A_{s}$ for $m=0.5,$ $f=250$ kHz, and two different
driving functions $V_{s}(t)$, namely Eq. (\ref{vs1}) for panel (a) and Eq. (\ref{vs2}) for panel (b).}
\label{single}%
\end{figure}

Before investigating the generation
of DBs in the electrical lattice, we have firstly examined in detail the response of an
isolated circuit cell. Figure \ref{single} shows the typical
response of such a circuit cell, namely a nonlinear resonance
curve over a certain range of values of the amplitude $A_{s}$ wherein two
different periodic attractors coexist, one of them exhibiting small-amplitude
oscillations whereas the other one features large-amplitude oscillations. Note that the
coexistence of these two periodic attractors with clearly different amplitudes
is a key ingredient for the existence of DBs \cite{breather_reviews}.

\begin{figure}[h]
\includegraphics[width=0.4\textwidth]{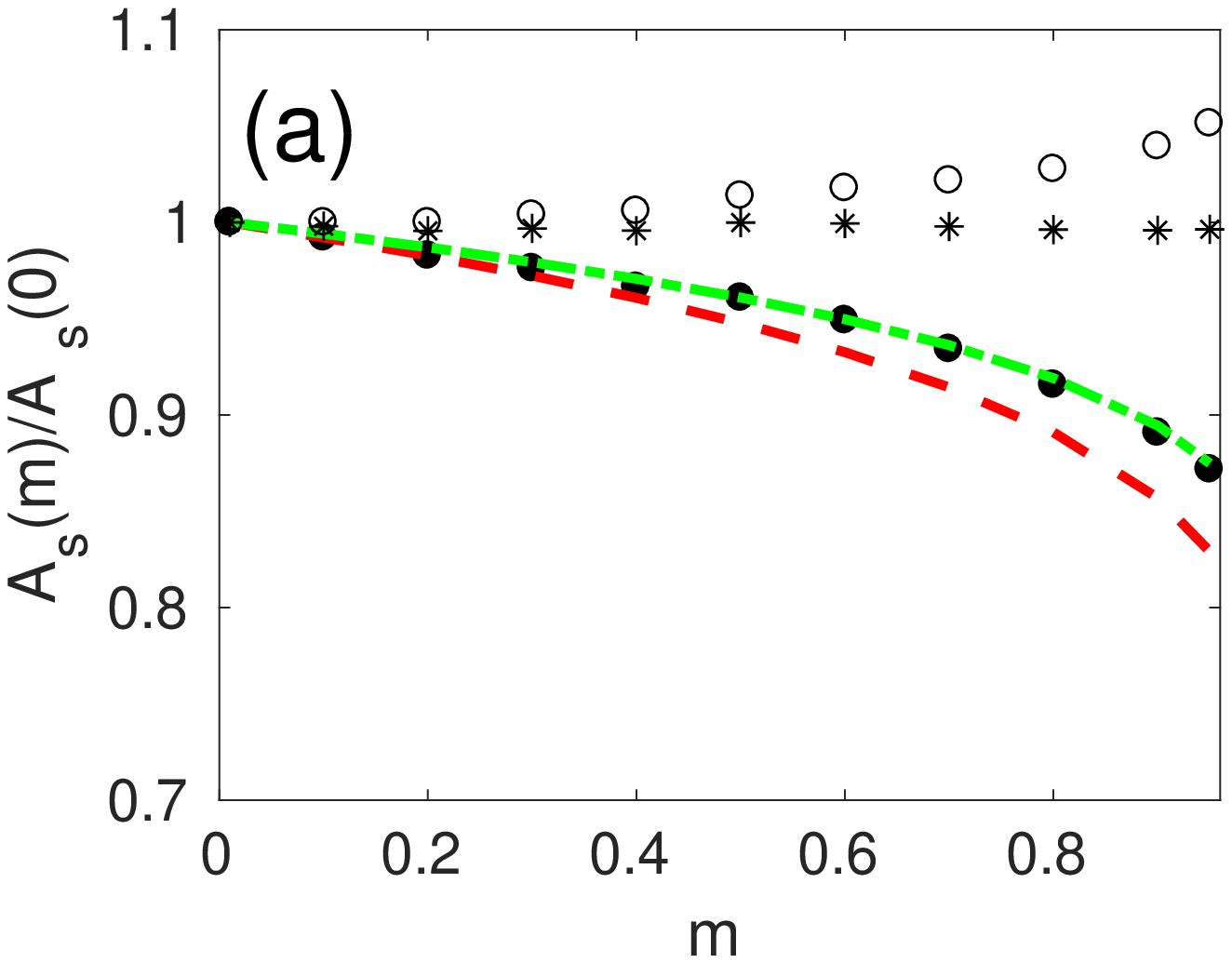}
\includegraphics[width=0.4\textwidth]{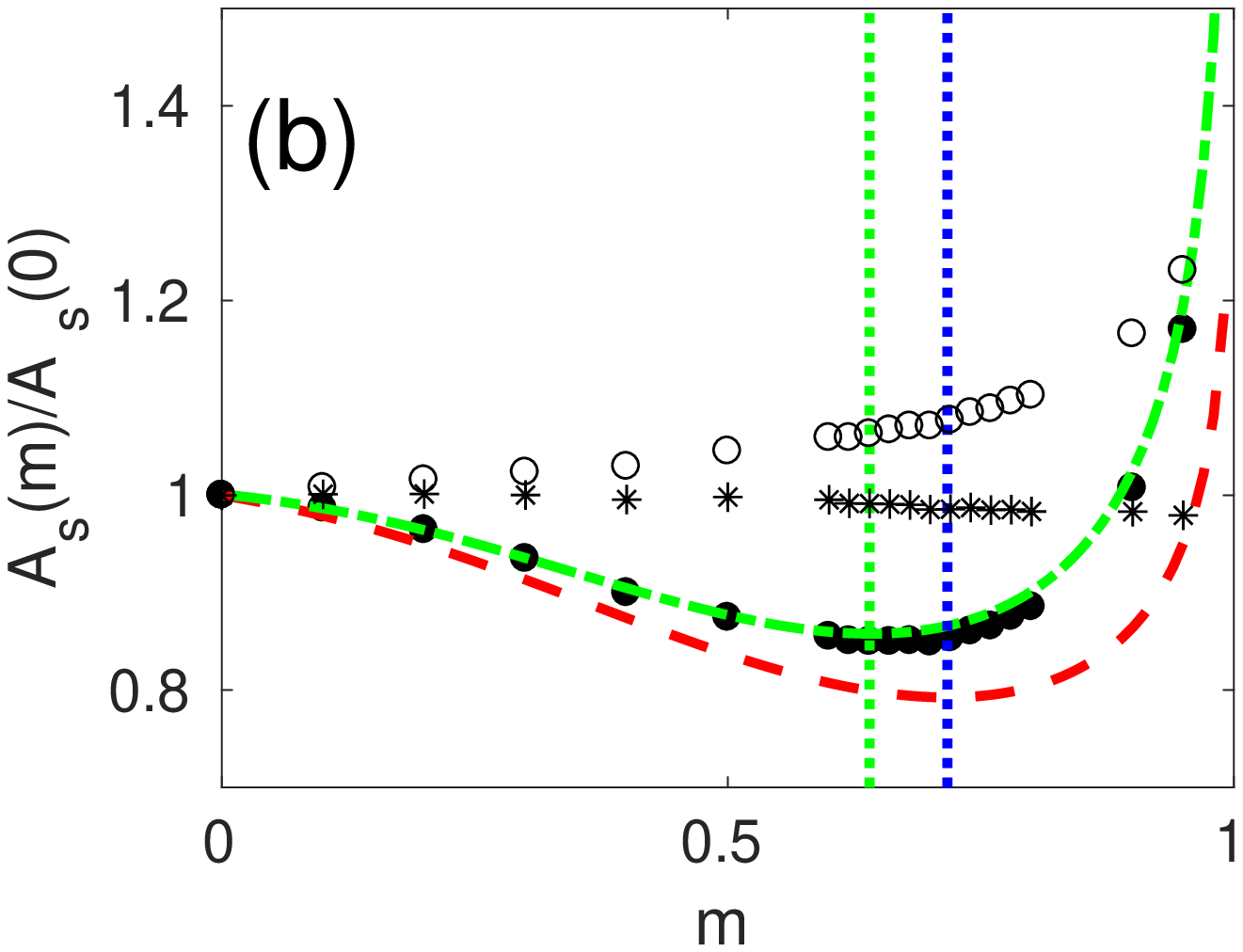}
\includegraphics[width=0.4\textwidth]{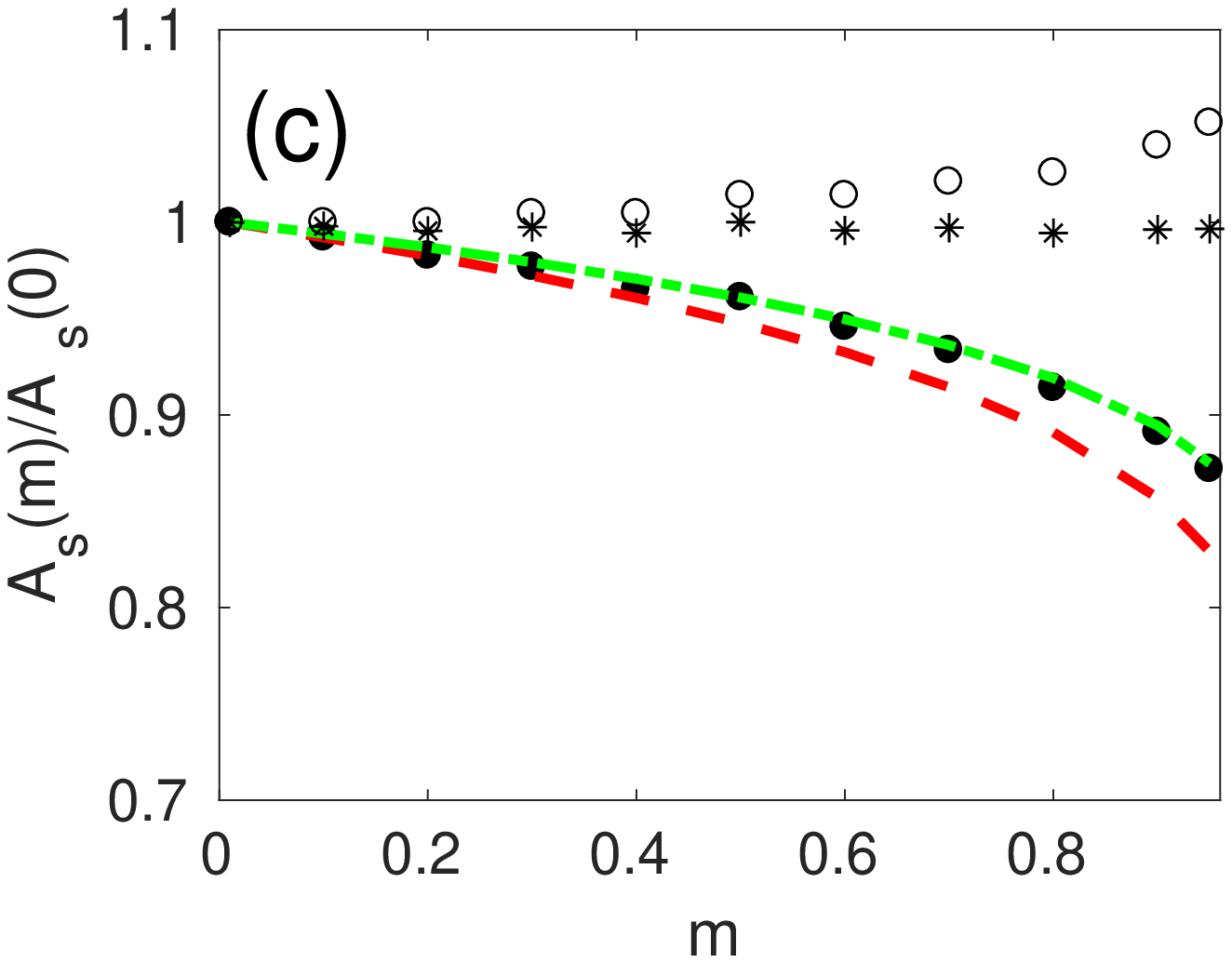}
\includegraphics[width=0.4\textwidth]{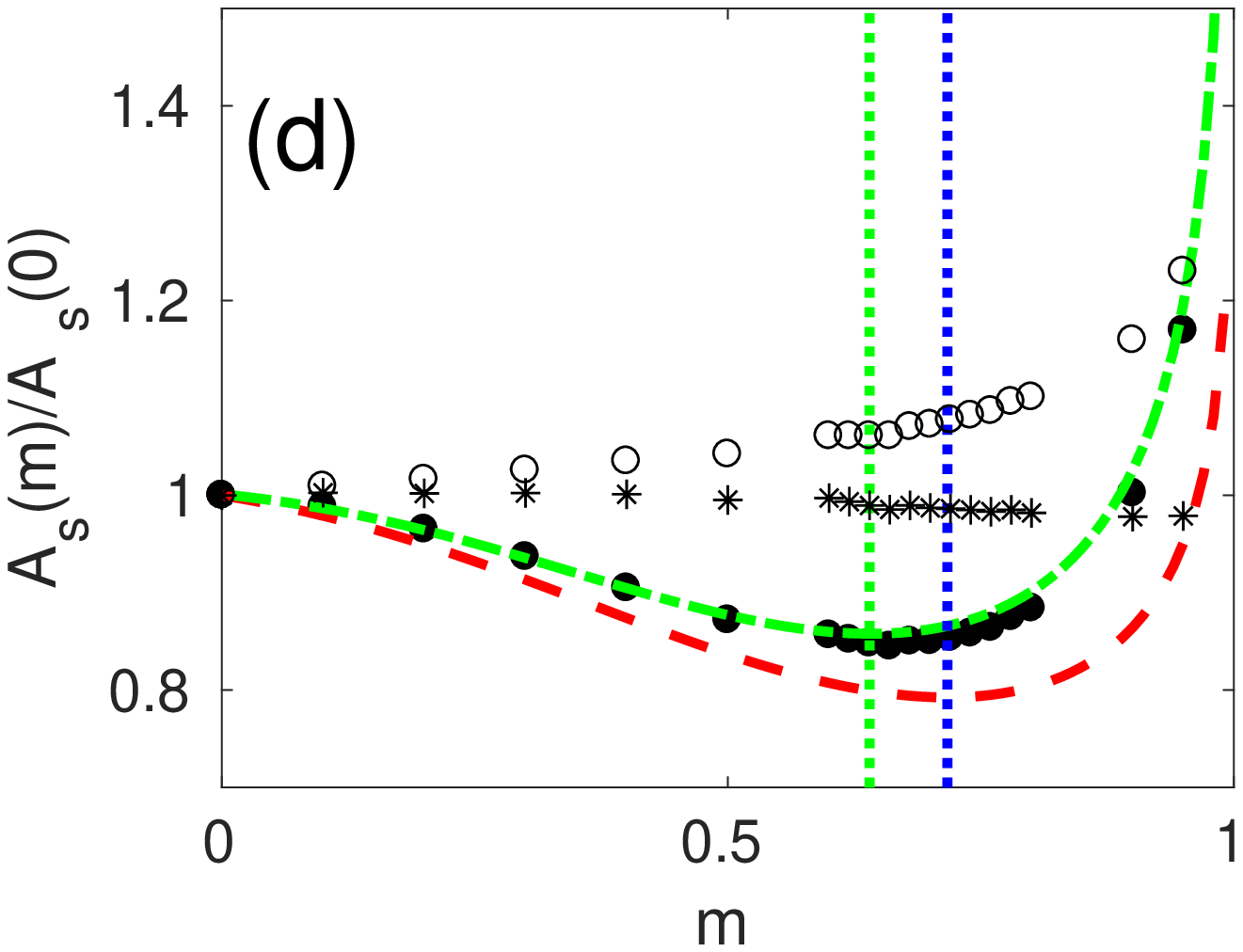}
\caption{(Color online) Critical values
of the amplitude $A_{s}$ giving rise to stable periodic attractors as a
function of the shape parameter $m$ for $f=250$ kHz. (a, b) Maximum values of
$A_{s}$ for which stable small-amplitude oscillations exist. (c, d) Minimum
values of $A_{s}$ for which stable large-amplitude oscillations exist. Dots represent experimental data, dashed (red) lines
represent the inverse of the (normalized) impulses $I^{(1,2)}(m=0)/I^{(1,2)}(m)$, dash-dot (green) lines represent
the inverse of the (normalized) first Fourier coefficient $G^{(1,2)}_1(m=0)/G^{(1,2)}_1(m)$, vertical (green) dotted lines indicate the value $m\simeq0.64$
and vertical (blue) dotted lines indicate the value $m\simeq0.717$. Panels
(a, c) correspond to the driving function (\ref{vs1}), while panels (b, d)
correspond to the driving function (\ref{vs2}). Note that, in (b) and (d) case, close to minimum we have taken more experimental measures
in order to improve the visualization of the sought phenomena. Open circles represent the experimental normalized amplitude $A_{s}/[I^{(1,2)}(m=0)/I^{(1,2)}(m)]$ and 
asterisks the corresponding experimental normalized amplitude $A_{s}/[G^{(1,2)}_1(m=0)/G^{(1,2)}_1(m)]$.}%
\label{fig_cri}%
\end{figure}

Figure \ref{fig_cri} shows the critical values of the amplitude
$A_{s}$ giving rise to stable small-amplitude and large-amplitude periodic
attractors as a function of the shape parameter $m$ (maximum and minimum
values of $A_{s}$, respectively). For the driving function (\ref{vs1}), we
have found that the threshold amplitude exhibits a monotonically decreasing behavior
as a function of the shape parameter, as expected from the monotonically
increasing behavior of its impulse. For the driving function (\ref{vs2}), we
found that the threshold amplitude follows the inverse behavior of its impulse
such that there exists a minimum threshold at a critical value of the shape
parameter: $m=m_{c}\approx0.64$. Note that this critical value is relatively close
to the value $m=m_{max}\approx0.717$ at which the impulse presents a single
maximum. Specifically, the critical value $m_{c}\approx0.64$ corresponds to the value of $m$ where the first harmonic of
the Fourier expansion of the PE (\ref{g2}) presents a single maximum.
In general, we have found that the threshold 
follows the inverse normalized 
first Fourier coefficients, 
which is in agreement with recently obtained numerical results \cite{CCP16} and is explicitly confirmed by our experiments (asterisks in Figs. 6 and 8).

\begin{figure}[h]
\includegraphics[width=0.5\textwidth]{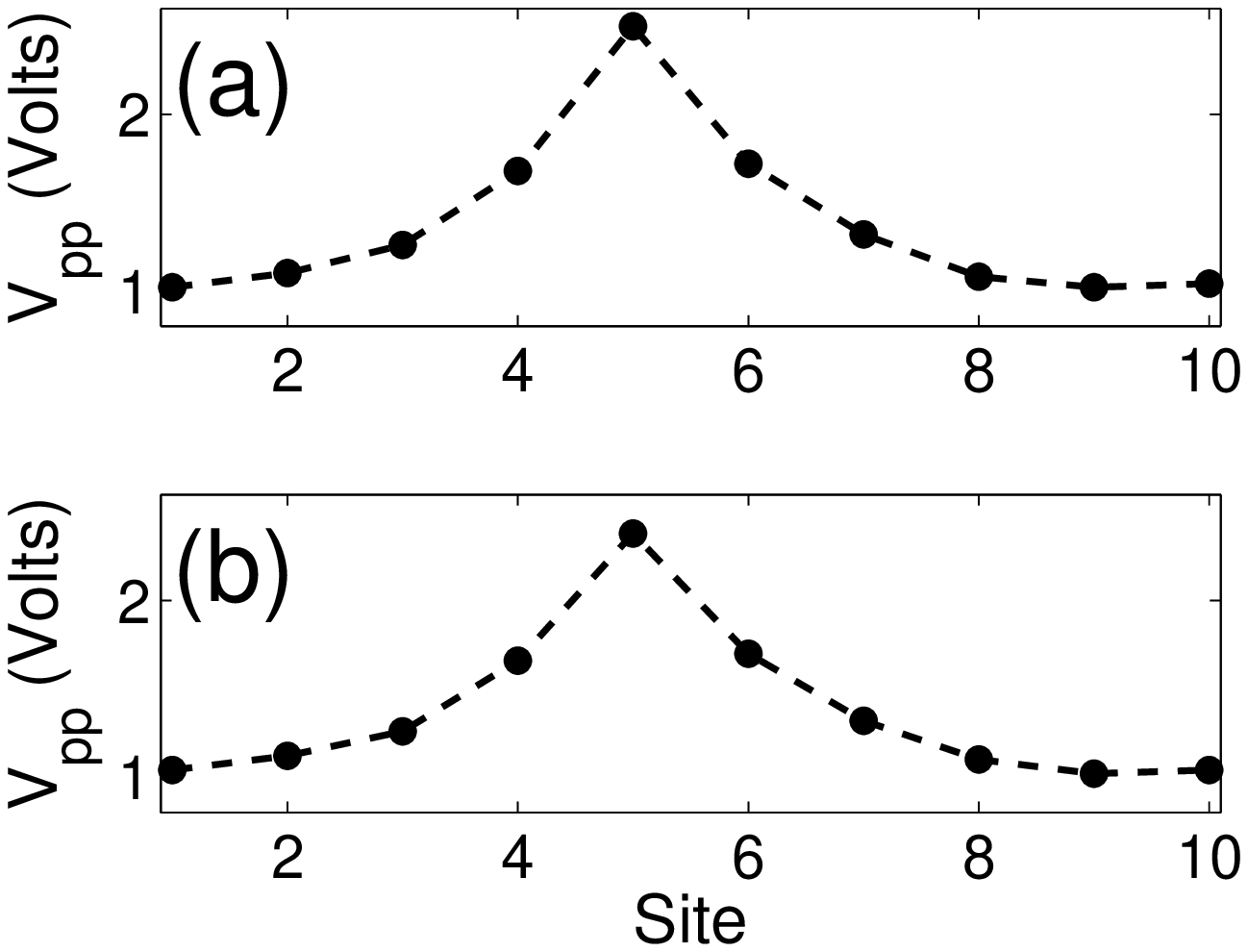}
\caption{Experimental (solid circles) breather profiles
for $f=250$ kHz (peak-to-peak value) and $m=0.95$.  (a) corresponds to the case (\ref{vs1}), where $A_{s}=12.4$ Volts
, while (b) corresponds to the case (\ref{vs2}), where $A_{s}=16.7$ Volts. Values of $A_{s}$ correspond to a threshold situation, because
of smaller values of $A_{s}$ implies that breather disappears.}%
\label{breather}%
\end{figure}

Next, we discuss the impulse-induced generation of DBs in the full
electrical line. After fixing the frequency to 250 kHz, we have found that, as expected, there exists a
minimum threshold value of the amplitude $A_{s}$ for which a stable one-peak
localized excitation emerges, with a typical profile as in the
example shown in Fig. \ref{breather}. In general, as the amplitude $A_{s}$ is
increased, stable multi-peak breathers emerge and the DBs scenario becomes
ever more complex. The ranges of
amplitudes $A_{s}$ wherein there exist breathers having a different number of
peaks overlap with those ranges wherein there exist different families of
breathers having the same number of peaks ---a feature which seems to depend
sensitively on the existence of small lattice impurities in our (trading)
experimental components \cite{electric1}.

\begin{figure}[h]
\includegraphics[width=0.4\textwidth]{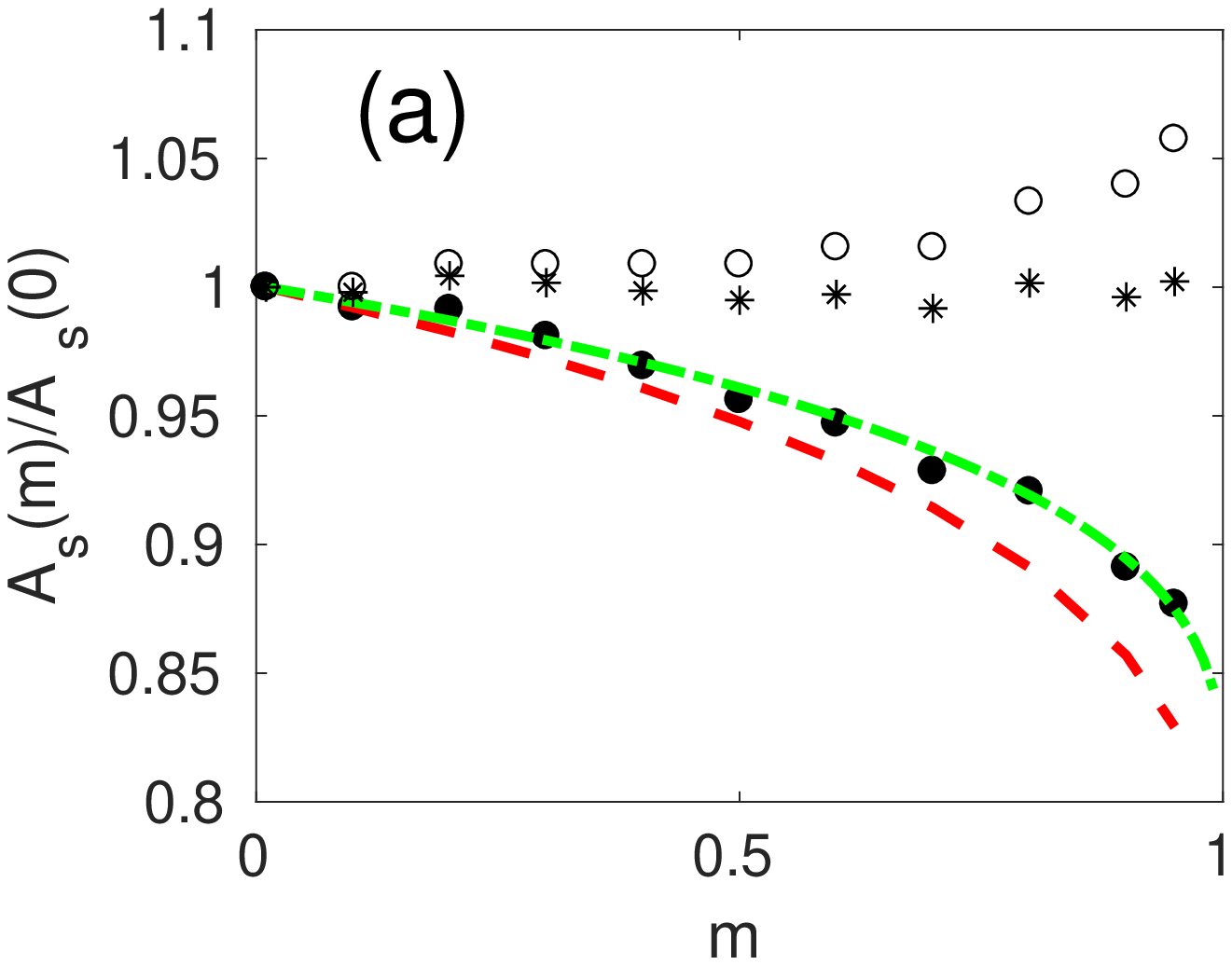}
\includegraphics[width=0.4\textwidth]{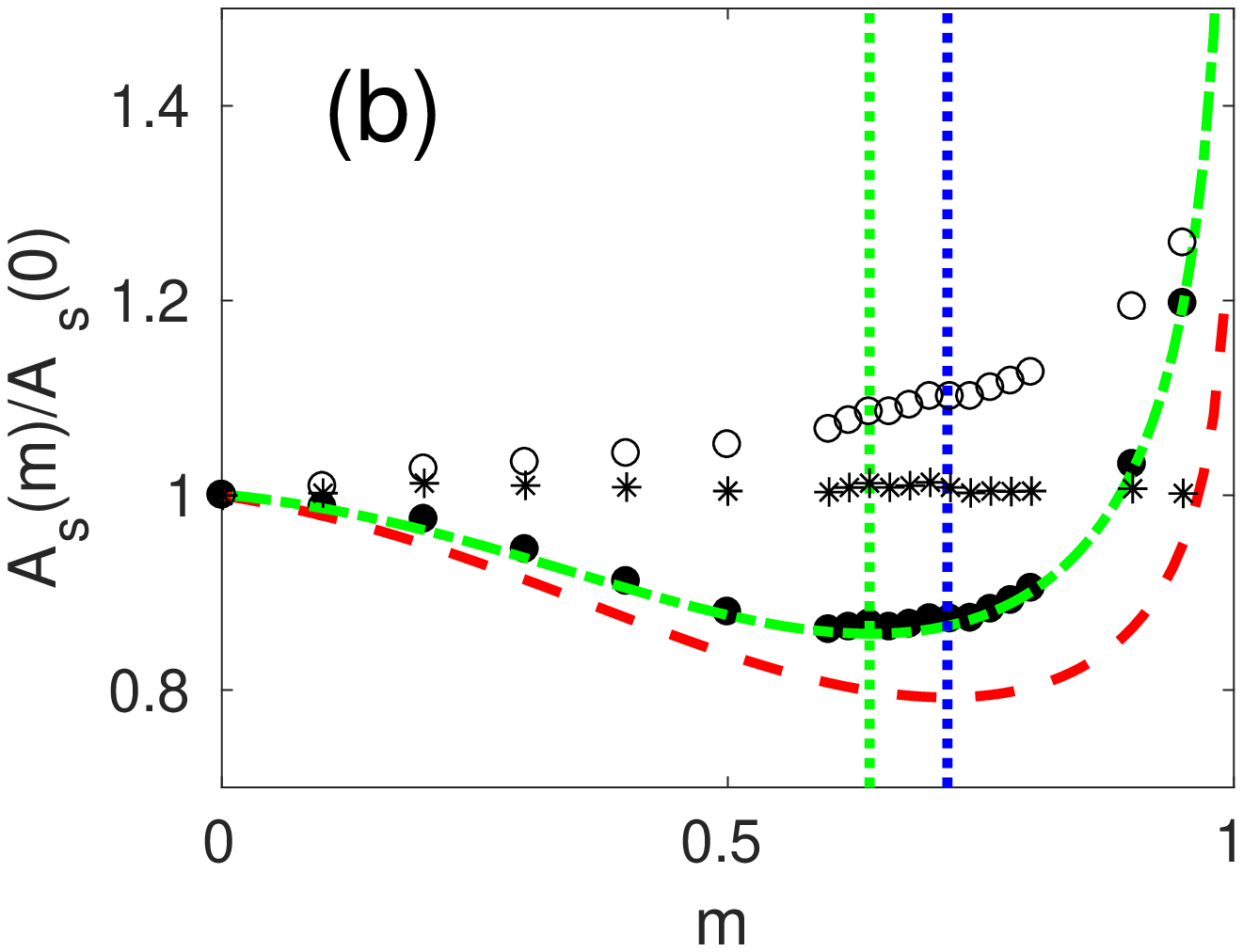}
\caption{(Color online) Minimum value of the amplitude $A_{s}$ giving rise to the existence of a stable discrete
breathers as a function of the shape parameter $m$ for $f=250$ kHz. Dots represent experimental values, 
dashed (red) lines
represent the inverse of the (normalized) impulses $I^{(1,2)}(m=0)/I^{(1,2)}(m)$, dash-dot (green) lines represent
the inverse of the (normalized) first Fourier coefficient $G^{(1,2)}_1(m=0)/G^{(1,2)}_1(m)$, vertical (green) dotted lines indicate the value $m\simeq0.64$
and vertical (blue) dotted lines indicate the value $m\simeq0.717$.  Panels (a) and (b) correspond to the driving 
functions (\ref{vs1}) and (\ref{vs2}), respectively. Open circles represent the experimental normalized amplitude $A_{s}/[I^{(1,2)}(m=0)/I^{(1,2)}(m)]$ and 
asterisks the corresponding experimental normalized amplitude $A_{s}/[G^{(1,2)}_1(m=0)/G^{(1,2)}_1(m)]$.Note that, in panel (b), we have taken more experimental
measures close to minimum for the sake of a better visualization the phenomenon. }%
\label{breather_c}%
\end{figure}

Figure \ref{breather_c} shows an illustrative instance
of the effect of the transmitted impulse in the generation of stable DBs. We
can see once again that, for the driving function (\ref{vs1}), the
threshold amplitude exhibits a monotonically decreasing behavior as a function
of the shape parameter, whereas for the driving function (\ref{vs2}) we found
that the threshold amplitude follows the inverse behavior of its impulse, such
that there exists a minimum threshold at a critical value of the shape
parameter: $m=m_{c}\approx0.64$, which is relatively close to the value
$m=m_{max}\approx0.717$ at which the impulse presents a single maximum in the
aforementioned sense. Note that the first Fourier coefficient of the driving
function (7) presents a single minimum at $m=m_{c}\approx0.64$. The proximity
of the values $m_{c}\approx0.64$ and $m_{max}\approx0.717$ can be understood
from the fact that the first Fourier coefficient of the driving function (7),
$G_{1}(m)$, already contains almost all the information of (7) regarding the
effect of its impulse, which is in turn a consequence of the extremely rapid
convergence of its Fourier expansion even for $m$ values very close to 1,
this being ultimately due to the dependence of $K(m)$ on $m$ \cite{17}.

\section{Conclusions} We have experimentally investigated the effectiveness of generic periodic
excitations of variable waveform at generating discrete breathers in
one-dimensional nonlinear electrical lattices. Specifically, we have
experimentally demonstrated for the first time that the impulse transmitted by
generic periodic excitations is a fundamental quantity providing a complete
control of their effectiveness at generating discrete breathers in real-world
electrical lines capable of presenting these intrinsic localized modes. We
have analytically shown that this effectiveness is due to a correlation
between increases in the transmitted impulse and increases in energy. Future
work may extend the present impulse-induced breather-generation scenario to
optimize the generation and control of diverse nonlinear localized
excitations, such as kinks, solitons and vortices.

\begin{acknowledgments}
R.C. gratefully acknowledges financial support from the Ministerio de
Econom\'{\i}a y Competitividad (MINECO, Spain) through Project No.
FIS2012-34902 cofinanced by FEDER funds, and from the Junta de Extremadura
(JEx, Spain) through Project No. GR15146. J.C.-M. thanks financial support from MAT2016-79866-R
project (AEI/FEDER, UE).
Authors thank to A.
Carretero-Ch\'avez, F. Palmero-Ramos and Francisco J. Vega
Narv\'aez for their help in the experimental setup. F. P.
acknowledges Dickinson College for hospitality and support.
\end{acknowledgments}

\end{document}